\documentclass[aps,prl,twocolumn,superscriptaddress,letterpaper]{revtex4}

\usepackage{graphicx}% Include figure files
\usepackage{dcolumn}% Align table columns on decimal point
\usepackage{bm}% bold math
\usepackage{amsmath, amssymb}% 
\usepackage{epstopdf}
\usepackage{color}
\usepackage[titletoc,toc,title]{appendix}
\usepackage{lipsum}

\usepackage{hhline}
\usepackage{multirow}
\usepackage{float}
\floatstyle{plaintop}
\restylefloat{table}
\newcommand{\eepp}{\mbox{($e,e'pp$)~}}
\newcommand{\eep}{\mbox{($e,e'p$)~}} 
\newcommand{\een}{\mbox{($e,e'n$)~}}

\newcommand{\eenp}{\mbox{$(e,e'np)$~}}
\newcommand{\eepn}{\mbox{$(e,e'pn)$~}}

\newcommand{\Aeepp}{\mbox{$A(e,e'pp)$~}}
\newcommand{\Aeep}{\mbox{$A(e,e'p)$~}}

\newcommand{\Aeenp}{\mbox{$A(e,e'np)$~}}
\newcommand{\Aeepn}{\mbox{$A(e,e'pn)$~}} 
\newcommand{\Apppn}{\mbox{$A(p,2pn)$~}} 

\usepackage[]{changes} % final
\begin{document}
%\linenumbers

\title{Direct Observation of Proton-Neutron Short-Range Correlation Dominance in Heavy Nuclei}

%%%%%%%%%%%%%%%%%%%%%%%%%%%%%%%%%%%%%%%%%%%% 
\newcommand*{\TAU }{School of Physics and Astronomy, Tel Aviv University, Tel Aviv 69978, Israel}
\newcommand*{\TAUindex}{1}
\affiliation{\TAU} 
\newcommand*{\MIT }{Massachusetts Institute of Technology, Cambridge, Massachusetts 02139, USA}
\newcommand*{\MITindex}{2}
\affiliation{\MIT} 
\newcommand*{\HUJI}{The Racah Institute of Physics, The Hebrew University, Jerusalem, Israe}
\newcommand*{\HUJIindex}{3}
\affiliation{\HUJI} 
\newcommand*{\ODU}{Old Dominion University, Norfolk, Virginia 23529}
\newcommand*{\ODUindex}{4}
\affiliation{\ODU} 
\newcommand*{\UTFSM}{Universidad T\'{e}cnica Federico Santa Mar\'{i}a, Casilla 110-V Valpara\'{i}so, Chile}
\newcommand*{\UTFSMindex}{5}
\affiliation{\UTFSM}
\newcommand*{\ANL}{Argonne National Laboratory, Argonne, Illinois 60439}
\newcommand*{\ANLindex}{5}
\affiliation{\ANL}
\newcommand*{\ASU}{Arizona State University, Tempe, Arizona 85287-1504}
\newcommand*{\ASUindex}{6}
\affiliation{\ASU}
\newcommand*{\CANISIUS}{Canisius College, Buffalo, NY}
\newcommand*{\CANISIUSindex}{7}
\affiliation{\CANISIUS}
\newcommand*{\CMU}{Carnegie Mellon University, Pittsburgh, Pennsylvania 15213}
\newcommand*{\CMUindex}{8}
\affiliation{\CMU}
\newcommand*{\CUA}{Catholic University of America, Washington, D.C. 20064}
\newcommand*{\CUAindex}{9}
\affiliation{\CUA}
\newcommand*{\SACLAY}{IRFU, CEA, Universit'e Paris-Saclay, F-91191 Gif-sur-Yvette, France}
\newcommand*{\SACLAYindex}{10}
\affiliation{\SACLAY}
\newcommand*{\CNU}{Christopher Newport University, Newport News, Virginia 23606}
\newcommand*{\CNUindex}{11}
\affiliation{\CNU}
\newcommand*{\UCONN}{University of Connecticut, Storrs, Connecticut 06269}
\newcommand*{\UCONNindex}{12}
\affiliation{\UCONN}
\newcommand*{\DUKE}{Duke University, Durham, North Carolina 27708-0305}
\newcommand*{\DUKEindex}{13}
\affiliation{\DUKE}
\newcommand*{\FU}{Fairfield University, Fairfield CT 06824}
\newcommand*{\FUindex}{14}
\affiliation{\FU}
\newcommand*{\FERRARAU}{Universita' di Ferrara , 44121 Ferrara, Italy}
\newcommand*{\FERRARAUindex}{15}
\affiliation{\FERRARAU}
\newcommand*{\FIU}{Florida International University, Miami, Florida 33199}
\newcommand*{\FIUindex}{16}
\affiliation{\FIU}
\newcommand*{\FSU}{Florida State University, Tallahassee, Florida 32306}
\newcommand*{\FSUindex}{17}
\affiliation{\FSU}
\newcommand*{\GWUI}{The George Washington University, Washington, DC 20052}
\newcommand*{\GWUIindex}{18}
\affiliation{\GWUI}
\newcommand*{\ISU}{Idaho State University, Pocatello, Idaho 83209}
\newcommand*{\ISUindex}{19}
\affiliation{\ISU}
\newcommand*{\INFNFE}{INFN, Sezione di Ferrara, 44100 Ferrara, Italy}
\newcommand*{\INFNFEindex}{20}
\affiliation{\INFNFE}
\newcommand*{\INFNFR}{INFN, Laboratori Nazionali di Frascati, 00044 Frascati, Italy}
\newcommand*{\INFNFRindex}{21}
\affiliation{\INFNFR}
\newcommand*{\INFNGE}{INFN, Sezione di Genova, 16146 Genova, Italy}
\newcommand*{\INFNGEindex}{22}
\affiliation{\INFNGE}
\newcommand*{\INFNRO}{INFN, Sezione di Roma Tor Vergata, 00133 Rome, Italy}
\newcommand*{\INFNROindex}{23}
\affiliation{\INFNRO}
\newcommand*{\INFNTUR}{INFN, Sezione di Torino, 10125 Torino, Italy}
\newcommand*{\INFNTURindex}{24}
\affiliation{\INFNTUR}
\newcommand*{\ORSAY}{Institut de Physique Nucl'eaire, IN2P3-CNRS, Universit'e Paris-Sud, Universit'e Paris-Saclay, F-91406 Orsay, France}
\newcommand*{\ORSAYindex}{25}
\affiliation{\ORSAY}
\newcommand*{\ITEP}{Institute of Theoretical and Experimental Physics, Moscow, 117259, Russia}
\newcommand*{\ITEPindex}{26}
\affiliation{\ITEP}
\newcommand*{\JMU}{James Madison University, Harrisonburg, Virginia 22807}
\newcommand*{\JMUindex}{27}
\affiliation{\JMU}
\newcommand*{\KNU}{Kyungpook National University, Daegu 41566, Republic of Korea}
\newcommand*{\KNUindex}{28}
\affiliation{\KNU}
\newcommand*{\LAMAR}{Lamar University, 4400 MLK Blvd, PO Box 10009, Beaumont, Texas 77710}
\newcommand*{\LAMARindex}{29}
\affiliation{\LAMAR}
\newcommand*{\MISS}{Mississippi State University, Mississippi State, MS 39762-5167}
\newcommand*{\MISSindex}{30}
\affiliation{\MISS}
\newcommand*{\UNH}{University of New Hampshire, Durham, New Hampshire 03824-3568}
\newcommand*{\UNHindex}{31}
\affiliation{\UNH}
\newcommand*{\NSU}{Norfolk State University, Norfolk, Virginia 23504}
\newcommand*{\NSUindex}{32}
\affiliation{\NSU}
\newcommand*{\OHIOU}{Ohio University, Athens, Ohio 45701}
\newcommand*{\OHIOUindex}{33}
\affiliation{\OHIOU}

\newcommand*{\RPI}{Rensselaer Polytechnic Institute, Troy, New York 12180-3590}
\newcommand*{\RPIindex}{34}
\affiliation{\RPI}
\newcommand*{\URICH}{University of Richmond, Richmond, Virginia 23173}
\newcommand*{\URICHindex}{35}
\affiliation{\URICH}
\newcommand*{\ROMAII}{Universita' di Roma Tor Vergata, 00133 Rome Italy}
\newcommand*{\ROMAIIindex}{36}
\affiliation{\ROMAII}
\newcommand*{\MSU}{Skobeltsyn Institute of Nuclear Physics, Lomonosov Moscow State University, 119234 Moscow, Russia}
\newcommand*{\MSUindex}{37}
\affiliation{\MSU}
\newcommand*{\SCAROLINA}{University of South Carolina, Columbia, South Carolina 29208}
\newcommand*{\SCAROLINAindex}{38}
\affiliation{\SCAROLINA}
\newcommand*{\TEMPLE}{Temple University, Philadelphia, PA 19122 }
\newcommand*{\TEMPLEindex}{39}
\affiliation{\TEMPLE}
\newcommand*{\JLAB}{Thomas Jefferson National Accelerator Facility, Newport News, Virginia 23606}
\newcommand*{\JLABindex}{40}
\affiliation{\JLAB}
\newcommand*{\GLASGOW}{University of Glasgow, Glasgow G12 8QQ, United Kingdom}
\newcommand*{\GLASGOWindex}{41}
\affiliation{\GLASGOW}
\newcommand*{\YORK}{University of York, York YO10, United Kingdom}
\newcommand*{\YORKindex}{42}
\affiliation{\YORK}
\newcommand*{\VIRGINIA}{University of Virginia, Charlottesville, Virginia 22901}
\newcommand*{\VIRGINIAindex}{43}
\affiliation{\VIRGINIA}
\newcommand*{\WM}{College of William and Mary, Williamsburg, Virginia 23187-8795}
\newcommand*{\WMindex}{44}
\affiliation{\WM}
\newcommand*{\YEREVAN}{Yerevan Physics Institute, 375036 Yerevan, Armenia}
\newcommand*{\YEREVANindex}{45}
\affiliation{\YEREVAN}
\newcommand*{\Genova}{Universita di Genova, Dipartimento di Fisica, 16146 Genova, Italy}
\newcommand*{\Genovaindex}{46}
\affiliation{\Genova}
\newcommand*{\NRCN}{Nuclear Research Centre Negev, Beer-Sheva, Israel}
\newcommand*{\NRCNindex}{47}
\affiliation{\NRCN}
\newcommand*{\NOWISU}{Idaho State University, Pocatello, Idaho 83209}
\newcommand*{\NOWJLAB}{Thomas Jefferson National Accelerator Facility, Newport News, Virginia 23606}

%%%%%%%%%%%%%%%%%%%% authors %%%%%%%%% 
\author{M.~Duer}
\affiliation{\TAU}
\author{A.~Schmidt}
\affiliation{\MIT}
\author{J.R. Pybus}
\affiliation{\MIT}
\author{E.P. Segarra}
\affiliation{\MIT}
\author{A.W. Denniston}
\affiliation{\MIT}
\author{R. Weiss}
\affiliation{\HUJI}
\author{O.~Hen}
\email[Contact Author \ ]{hen@mit.edu}
\affiliation{\MIT}
\author{E.~Piasetzky}
\affiliation{\TAU}
\author{L.B.~Weinstein}
\affiliation{\ODU}
\author{N. Barnea}
\affiliation{\HUJI}
% ...
\collaboration{The CLAS Collaboration}
\noaffiliation

\begin{abstract}

We measured the triple coincidence $A\eenp$ and $A\eepp$ reactions
on carbon, aluminum, iron, and lead targets at $Q^2 >$ 1.5
(GeV/c)$^2$, $x_B >$ 1.1 and missing momentum $> 400$ MeV/c. This
was the first direct measurement of both proton-proton ($pp$) and
neutron-proton ($np$) short-range correlated (SRC) pair knockout
from heavy asymmetric nuclei. For all measured nuclei, the average
proton-proton ($pp$) to neutron-proton ($np$) reduced cross-section
ratio is about $6\%$, in agreement with previous indirect
measurements. Correcting for Single-Charge Exchange effects
decreased the SRC pairs ratio to $\sim 3\%$, which is lower than
previous results. Comparisons to theoretical Generalized Contact
Formalism (GCF) cross-section calculations show good agreement using
both phenomenological and chiral nucleon-nucleon potentials, 
favoring a lower $pp$ to $np$ pair ratio. 
The ability of the GCF calculation to describe the experimental data using either 
phenomenological or chiral potentials suggests possible reduction of scale- and 
scheme-dependence in cross section ratios.
Our results also support the high-resolution description of high-momentum states being
predominantly due to nucleons in SRC pairs.

\end{abstract}

\maketitle

% =====================================================================

%intro
Recent high-momentum transfer measurements have shown that nucleons in
the nuclear ground state can form temporary pairs with large relative
momentum and small center-of-mass (CM) momentum~\cite{Hen:2016kwk,Atti:2015eda}. 
These pairs are referred to as short range correlated (SRC) pairs. 
The formation of SRC pairs in heavy, asymmetric nuclei has implications for momentum 
sharing between protons and neutrons in these nuclei~\cite{hen14, Sargsian:2012sm, ryckebusch15, duer18, Ryckebusch:2018rct}, 
our understanding of the properties of very asymmetric cold dense nuclear systems such as neutron
stars~\cite{frankfurt08b, hen15, Li:2018lpy}, and the relative modification of proton and neutron structure 
in nuclei (the EMC effect)~\cite{Hen:2016kwk, weinstein11, hen11, Hen12, Hen:2013oha, Chen:2016bde, Ciofi07, Schmookler:2019nvf}.

Properties of SRC pairs are primarily inferred from measurements of
exclusive electron- and proton-induced triple-coincidence hard breakup
reactions. In these experiments, a nucleon is knocked out of the
nucleus via a high-momentum transfer reaction and detected in
coincidence with the scattered probe and a recoil nucleon balancing a
large missing momentum. Previous measurements of such \Aeepp, \Aeepn
and \Apppn reactions in light symmetric nuclei ($^4$He and $^{12}$C),
showed that neutron-proton ($np$) SRC pairs are nearly $20$ times as
prevalent as proton-proton ($pp$) pairs and, by inference,
neutron-neutron ($nn$) pairs~\cite{tang03,piasetzky06,subedi08,korover14}. This
$np$-pair dominance was explained as being due to the dominance of the tensor part 
of the nucleon-nucleon force at high relative momenta~\cite{schiavilla07, sargsian05, Alvioli:2007zz, neff15}.
See recent reviews in~\cite{Hen:2016kwk, Atti:2015eda}.

For nuclei heavier than carbon, the predominance of $np$-SRC pairs was never
extracted directly from measurements of the exclusive \Aeepp and
\Aeepn reactions. Instead, it was inferred from measurements of the
exclusive \Aeepp and semi-inclusive \Aeep reactions, by assuming that
all high missing-momentum nucleons knocked out in the \Aeep reaction
are part of SRC pairs \cite{hen14}. Thus, \Aeep events without a
correlated recoil proton were attributed to breakup of $np$-SRC pairs.

Here we report, for the first time, the simultaneous measurement of
exclusive triple coincidence \Aeenp and \Aeepp reactions on carbon,
aluminum, iron, and lead. The new data confirm the previously deduced
$np$-SRC dominance without the assumptions required by previous
analyses \cite{hen14}. We also show that the new data agree with
factorized Generalized Contact Formalism (GCF)
calculations~\cite{Weiss:2015mba,Weiss:2016obx,Weiss:2018tbu} using
both phenomenological and Chiral $NN$ interactions. The agreement
between this new measurement and both the previous results and the GCF
calculations, supports the high-resolution description of high-momentum
nucleons in nuclei as predominantly members of SRC pairs.

\begin{figure} [t]
\includegraphics[width=\columnwidth, height=5.6cm]{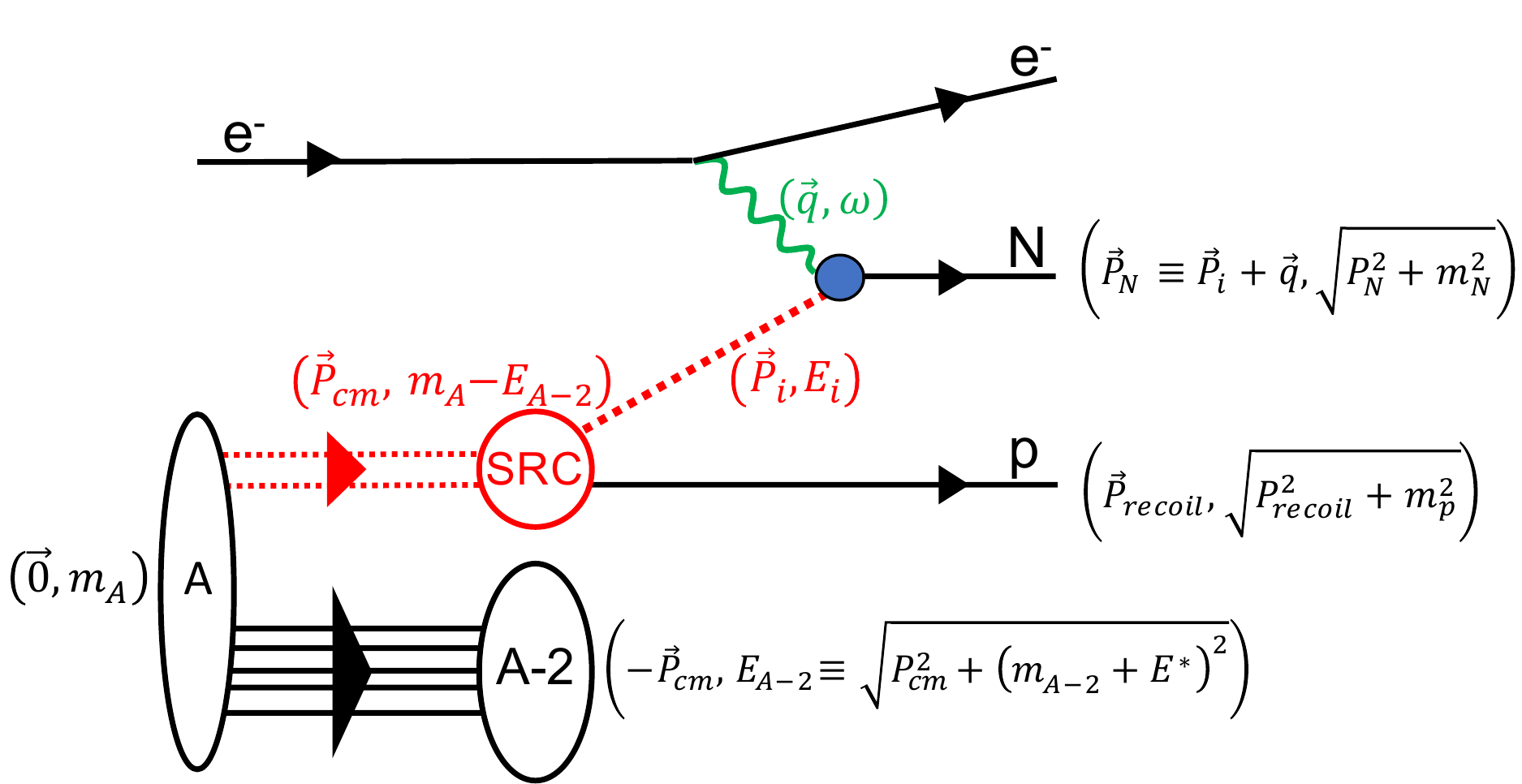}
\caption{\label{fig:reaction} (color online) Diagrammatic
representation and kinematics of the triple-coincidence $A(e,e'Np)$
reaction within the SRC breakup model. Dashed red lines represent
off-shell particles. Open ovals represent un-detected systems. Solid
black lines represent detected particles. The momentum and energy
of the particles is also indicated. See
text for details.}
\end{figure}

In the SRC description of high missing momentum nucleon knockout
reactions, the
nucleus is described as composed of an off-shell SRC pair (either $np$, $nn$ or $pp$)
with center of mass (total) momentum $\vec P_{cm}$ plus an on-shell $A-2$
residual system with momentum $-\vec P_{cm}$ (see Fig.~\ref{fig:reaction}). The incident
high-energy electron scatters from the nucleus by transferring a
single virtual photon, carrying momentum $\vec q$ and energy $\omega$,
to a single off-shell nucleon in the SRC-pair with initial momentum
$\vec{p}_i$ and energy $E_i$, a process we refer to as quasi-elastic
(QE) scattering. If that nucleon does not re-scatter as it leaves the
nucleus, it will emerge with momentum $\vec{p}_N = \vec{p}_i +
\vec{q}$. Thus, we can reconstruct the approximate initial momentum of
that nucleon from the measured missing momentum $\vec p_i \approx
\vec{p}_{miss} = \vec{p}_N - \vec{q}$. The correlated recoil proton
is an on-shell spectator that carries momentum $\vec{P}_{recoil} =
\vec{P}_{cm} - \vec{P}_{miss}$ and corresponding energy $E_{recoil} =
\sqrt{P_{recoil}^2 + m_p^2}$. The undetected residual $A-2$ system has
momentum $-\vec{P}_{cm}$ and can have excitation energy $E^*$.

%initial FSI discussion
SRC studies are typically done at $Q^2 =\vec q\thinspace^2 - \omega^2 > 1.5$
(GeV/c)$^2$, $x_B =Q^2/2m\omega >$ 1, (where $m$ is the nucleon mass)
anti-parallel kinematics, and missing momentum that exceeds the Fermi
momentum, i.e., $|\vec{p}_{miss}| > 300$ MeV/c
~\cite{hen14,korover14,duer18}. According to calculations, non-QE reaction
mechanisms (i.e., reactions other than the hard breakup of SRC pairs described above) are suppressed
under these
conditions~\cite{Frankfurt:1996xx,frankfurt08b,Atti:2015eda,Hen:2016kwk},
and the mechanism of Fig.~\ref{fig:reaction} should be a
valid description of the reaction.

Rescattering of the outgoing struck nucleon (final state interactions
or FSI) might alter the final state of the reaction. This
rescattering includes contributions from nuclear transparency (flux
reduction), small angle nucleon rescattering, and single-charge
exchange (SCX). However, these effects are significantly reduced in
cross-section ratios as compared to absolute
cross-sections~\cite{Hen:2016kwk, colle15, Colle:2015lyl, boeglin11}.
In addition, at the relevant high-$Q^2$, the cross-sections
approximately factorize and calculations of FSI, including both
outgoing-nucleon rescattering and SCX, are done using an Eikonal
approximation in a Glauber framework, which was shown to agree well
with data (see~\cite{Duer:2018sjb,hen12a,Atti:2015eda,Ryckebusch:2003fc,Colle:2015lyl,
Frankfurt:1996xx, Colle:2013nna, Dutta:2012ii, Frankfurt:2000ty,
Pandharipande:1992zz} and references therein). These calculations
show that small-angle rescattering (i.e., FSI that do not lead to a
reduction of flux) is largely confined to within the nucleons of the
pair~\cite{Frankfurt:1996xx,Sargsian:2001ax,Colle:2015lyl,
frankfurt08b}. Such re-scattering does not change the isospin
structure of SRC pairs, which is the focus of this analysis.

The theoretical description of high-momentum transfer measurements
presented above constitutes a valid simple reaction picture that is
consistent with both data and various ab-initio
calculations~\cite{Carlson:2014vla,Hen:2016kwk,Atti:2015eda}.
However, it is not the only possible description of such reactions.
Utilizing unitary freedom, one can always shift the complexities of
explicit two-body correlations from nuclear wave functions to the
interaction operators without changing the measured cross-sections
(i.e., shifting from a simple one-body operator acting on a
complicated ground state with SRC pairs, to a simple ground state with
complicated many-body interaction operators). Therefore, there is no
unique way to separate the description of the nuclear ground state
from that of the reaction mechanism (see, e.g., discussion in
Ref.~\cite{More17} for the deuteron photodisintegration case). 
For clarity of the discussion, this work focuses on a high-resolution reaction picture, 
using one-body operators. The results presented here can, however, be used to also 
constrain theoretical calculations done using the low-resolution many-body operators approach.

%%%%% event selection sensitivity table
\begin{table}[b]
\center
\caption{\label{tab:1} The $(e,e'Np)$ event selection cuts. Also shown is the
sensitivity of the $pp/np$ ratios to variations of the cuts. {*Both leading nucleon cuts were varied simultaneously.} }
\begin{tabular}{|c|c|c c c c|}
\hline
\multirow{2}{*}{Cut} &
\multicolumn{5}{c|}{Cut Sensitivity [$\%$]} \\
& Range & C & Al & Fe & Pb \\
\hline 
\hline
\footnotesize{$x_{B}>1.1$} & \footnotesize{$\pm$0.05} & \footnotesize{1.5} & \footnotesize{1.9} & \footnotesize{1.4} & \footnotesize{1.7}\\ 

\footnotesize{$0.62<|\vec{p}_N|/|\vec{q}|<1.1$} & \footnotesize{*$\pm$0.1} & \multirow{2}{*}{\footnotesize{2.7}} & \multirow{2}{*}{\footnotesize{2.5}} & \multirow{2}{*}{\footnotesize{2.3}} & \multirow{2}{*}{\footnotesize{2.2}}\\ 
\footnotesize{$\theta_{Nq}<25^{\circ}$} & \footnotesize{*$\pm5^{\circ}$} & & & &\\ 

\footnotesize{$m_{miss}<1.175$ GeV/c$^{2}$} & \footnotesize{$\pm$0.02 GeV/c$^{2}$} & \footnotesize{2.4} & \footnotesize{2.3} & \footnotesize{3.1} & \footnotesize{2.0}\\ 
\footnotesize{$0.4<p_{miss}<1$ GeV/c} & \footnotesize{$\pm$0.025 GeV/c} & \footnotesize{2.6} & \footnotesize{2.8} & \footnotesize{2.1} & \footnotesize{2.1}\\ 
\footnotesize{$p_{recoil}>0.35$ GeV/c} & \footnotesize{$\pm$0.025 GeV/c} & \footnotesize{2.4} & \footnotesize{2.6} & \footnotesize{2.3} & \footnotesize{2.7}\\ 
\footnotesize{SC Deposited Energy} & \footnotesize{cut ON/OFF} & \footnotesize{0.2} & \footnotesize{3.2} & \footnotesize{1.0} & \footnotesize{2.3}\\ 
\hline 
\footnotesize{Total} & & \footnotesize{5.3} & \footnotesize{6.3} & \footnotesize{5.2} & \footnotesize{5.4}\\ 
\hline 

\end{tabular}
\end{table}
%minimal description of Meytal's Nature
%The study of SRCs using semi-inclusive $A\eep$ and $A\een$ reactions, in the above-mentioned kinematics, was recently reported in~\cite{duer18}. 
%In that analysis, QE nucleon knockout events were divided into two kinematical regions, corresponding to electron scattering off high-missing-momentum ($|\vec{p}_{miss}| > k_F$) nucleons, presumably from an SRC pair, or from low-missing-momentum ($|\vec{p}_{miss}| < k_F$) nucleons, presumably from shell model states, where $k_F$ is the Fermi momentum of the nuclei. 
%Here we extend this study 
%of high missing-momentum semi-inclusive single-nucleon knockout events 
%to exclusive $A(e,e'np)$ and $A(e,e'pp)$ two-nucleon knockout events.
% by detecting a recoil proton in coincidence with the high-missing momentum $(e,e'p)$ and $(e,e'n)$ events.

%CLAS - generic
The data presented here were collected in 2004 in Hall B of the Thomas
Jefferson National Accelerator Facility (Jefferson Lab) in Virginia,
USA, and are re-analyzed here as part of the Jefferson Lab data-mining
initiative ~\cite{DataMining}. The experiment used a 5.01 GeV electron
beam incident on deuterium, carbon, aluminum, iron, and lead
targets~\cite{Hakobyan:2008zz}, and the CEBAF Large Acceptance
Spectrometer (CLAS)~\cite{Mecking:2003zu} to detect the scattered
electron, the knocked-out proton or neutron, and the recoil proton.

%CLAS - generic + PID
CLAS used a toroidal magnetic field and six independent sets of drift
chambers, time-of-flight (TOF) scintillation counters, Cherenkov
Counters (CC), and Electromagnetic Calorimeters (EC). Charged
particle momenta were inferred from their reconstructed trajectories
as they bend due to the influence of the toroidal magnetic
field. Electrons were identified by requiring a signal in the CC and
a characteristic energy deposition in the EC. Protons and
pions were identified by comparing their measured flight times and
momenta. For low-momentum particles ($p< 700$ MeV/c), proton/pion
separation was further improved by requiring the protons to deposit
more than $15$ MeV in the $5$-cm thick TOF counters. Neutrons were
identified by observing interactions in the forward EC (covering about
8$^\circ$ to 45$^\circ$) with no associated hit in the corresponding TOF
counter and no matching charged-particle track in the drift
chambers. The angle- and momentum-dependent neutron detection
efficiency and momentum reconstruction resolution were measured using
the exclusive $d(e,e'p\pi^+$$\pi^-)n$ and $d(e,e'p\pi^+\pi^-n)$
reactions. See the on-line supplemental information of
Refs.~\cite{hen14,duer18} for details of the analysis.

%%%%% pp/np ratios
\begin{table}[t]
\center
\caption{\label{tab:0} Measured $[A(e,e'pp)/2\sigma_{ep}] / [A(e,e'np)/\sigma_{en}]$ reduced cross-section ratios in percent units and their uncertainties divided into two recoil proton momentum bins. The first uncertainty is statistical while second is systematical. See text for details.}
\begin{tabular}{cccll}
\cline{1-3}
%\multicolumn{3}{|c|}{$\frac{\sigma_A(e,e'pp)/2\sigma_{ep}}{\sigma_A(e,e'np)/\sigma_{en}}$} & & \\ \cline{1-3}
\multicolumn{1}{|c|}{\multirow{2}{*}{A}} & \multicolumn{2}{c|}{$|P_{recoil}|$ [GeV/c]} & & \\ \cline{2-3}
\multicolumn{1}{|l|}{} & \multicolumn{1}{c|}{0.3 - 0.6} & \multicolumn{1}{c|}{0.6 - 1.0} & & \\ \cline{1-3}
\multicolumn{1}{|c|}{C} & \multicolumn{1}{c|}{5.33 $\pm$ 0.65 $\pm$ 0.35} & \multicolumn{1}{c|}{7.87 $\pm$ 1.68 $\pm$ 0.70} & & \\ \cline{1-3}
\multicolumn{1}{|c|}{Al} & \multicolumn{1}{c|}{5.33 $\pm$ 1.09 $\pm$ 0.33} & \multicolumn{1}{c|}{8.76 $\pm$ 3.63 $\pm$ 1.05} & & \\ \cline{1-3}
\multicolumn{1}{|c|}{Fe} & \multicolumn{1}{c|}{5.88 $\pm$ 0.68 $\pm$ 0.34} & \multicolumn{1}{c|}{6.53 $\pm$ 1.16 $\pm$ 0.41} & & \\ \cline{1-3}
\multicolumn{1}{|c|}{Pb} & \multicolumn{1}{c|}{5.71 $\pm$ 1.49 $\pm$ 0.39} & \multicolumn{1}{c|}{6.70 $\pm$ 1.93 $\pm$ 0.40} & & \\ \cline{1-3}
\multicolumn{1}{l}{} & \multicolumn{1}{l}{} & \multicolumn{1}{l}{} & & 
\end{tabular}
\end{table}
% event selection
We selected high missing-momentum \eep and \een events (i.e., events
with a ``leading'' proton or neutron) following the procedure of
Ref.~\cite{duer18} using the cuts detailed in Table~\ref{tab:1}. We
further required the detection of a lower-momentum recoil-proton ($350
\le |\vec{p}_{recoil}| \le 1000$ MeV/c) to obtain \eepp and \eenp
events. Since the recoil protons had relatively low momentum,
following~\cite{hen14} we corrected their momenta for energy loss in
the target and the CLAS detector.

% cross-section ratio extraction
As CLAS uses an open $(e,e')$ trigger, \Aeepp and \Aeenp reactions
were measured simultaneously. We matched the \Aeepp and \Aeenp
acceptances by considering only leading nucleons which were detected
in the phase-space region with good acceptance for both protons and
neutrons. To extract the \Aeepp / \Aeenp cross-section ratio from the
measured event yields, we weighted each event by the inverse of the
leading-nucleon detection efficiency. 

\begin{figure} [t]
\includegraphics[width=\columnwidth, height=5.8cm]{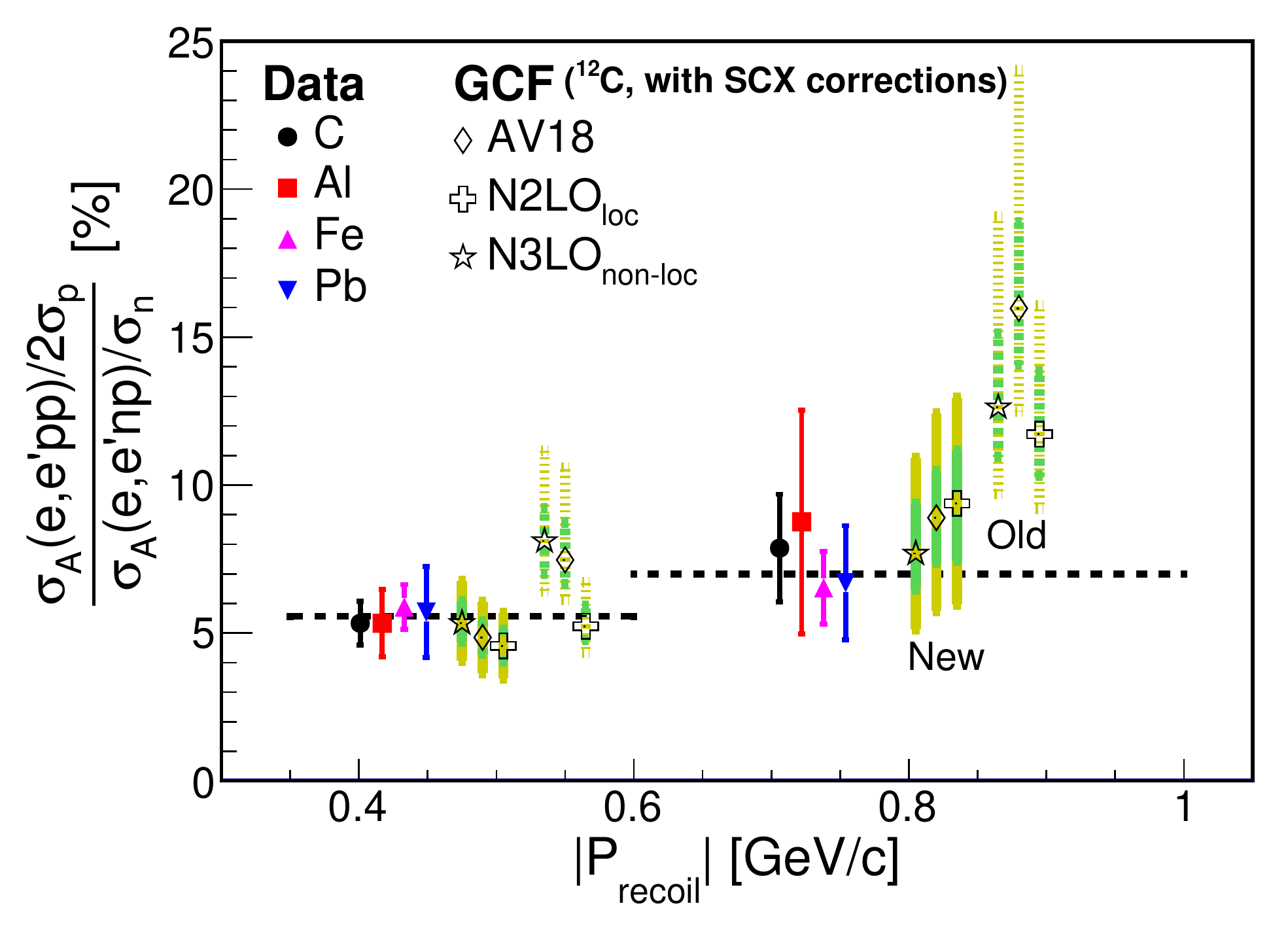}
\caption{\label{fig:0} (color online) Extracted reduced cross-section
ratios $R$ for $pp$ to $np$ SRC pair knockout as a function of recoil proton
momenta. Different filled symbols mark different nuclei. The black
dashed lines show the average cross-section ratio for all four
nuclei and their horizontal extents show the width of each recoil
proton momenta bin. The open symbols show the results of GCF
calculations for $^{12}C$ using three different $NN$ interactions. The inner (green)
and outer (yellow) bands represent the 68\% and 95\% confidence ranges of the
calculation. The points with dashed error bars correspond to GCF
calculations using the ``old'' $pp$ to $np$
contact ratios of Ref. \cite{Weiss:2018tbu} and the points with the
solid error bars use ``new'' contact ratios fit to this data set. See text for details.}
\end{figure}

Figure~\ref{fig:0} shows the resulting reduced cross-section ratio 
\begin{equation}
R= \frac{Y(A(e,e'pp)) / 2\sigma_{ep}}{Y(A(e,e'np)) / \sigma_{en}}
\label{eq:R}
\end{equation}
for all
measured nuclei (where $Y$ is the efficiency-corrected yield, and
$\sigma_{ep}$ and $\sigma_{en}$ are the elementary electron-proton and
electron-neutron cross sections, respectively~\cite{Ford:2014yua}),
divided into two bins of recoil proton momenta ($350 - 600$ and $600 -
1000$ MeV/c).  
The weighting
factors of $1/(2\sigma_{ep})$ and $1/\sigma_{en}$ were applied event-by-event to account for the
different elementary electron-nucleon cross sections and the
different nucleon counting. The error bars show both statistical and
systematical uncertainties added in quadrature. The latter include
sensitivity of the extracted cross-section ratio to the event
selection cuts detailed in Table~\ref{tab:1}, uncertainties in the
neutron and proton detection efficiencies, and a small difference for
the leading proton and neutron transparencies in iron and
lead~\cite{Colle:2015lyl, cosynPrivate} (see table I in the
online supplementary materials).
Numerical values for the extracted cross-section ratios are listed in Table~\ref{tab:0}.

The reduced cross-section ratio $R$ in each bin is
$A$-independent, and increases from an average of $5.5\pm0.4\%$ at the lower
$P_{recoil}$ bin to $7.0\pm0.9\%$ at the higher bin. Its small value
is consistent with $np$-SRC pairs being $15 - 20$ times more
abundant than $pp$-SRC pairs. The increase between the two
bins is also consistent with the expected increased contribution of
$pp$-SRC pairs at higher relative momenta where the tensor part of the
nuclear interaction is less predominant~\cite{korover14}.

In order to extract the ratio of $np$ to $pp$ pairs in the nucleus
from the reduced cross-section ratio, we need to correct for the
attenuation and SCX interactions (e.g., $(n,p)$ and $(p,n)$ reactions)
of the nucleons as they exit the nucleus. At the measured outgoing
nucleon momenta, the $pp$ and $nn$ elastic scattering cross-sections
are similar, so nucleon attenuation largely cancels in the \Aeenp /
$\Aeepp$ cross-section ratio (see~\cite{Colle:2015lyl} for details).
However, SCX can increase the observed reduced cross-section
ratio. Because there are so many more $np$- than $pp$-SRC pairs, $np$
pair knockout, followed by an $(n,p)$ charge-exchange reaction, could
comprise a large fraction of the measured \Aeepp events. Correcting
for this effect will decrease the extracted ratio of $pp$- to $np$-SRC pairs
relative to the measured reduced cross-section ratio $R$. Thus, $R$
is an upper limit on the $pp$- to $np$-SRC pairs ratio.

We calculated scattering cross sections for the reaction diagram shown
in Fig.~\ref{fig:reaction} using the factorized GCF model
\cite{Weiss:2018tbu}.  These GCF calculations use SRC-pair relative
momentum distributions calculated with a given $NN$ potential (which
are the same for all nuclei), the measured $P_{cm}$ distributions
\cite{Cohen:2018gzh}, and the relative abundances of $np$, $pp$, and
$nn$ pairs (i.e., the ``contacts'') in a given nucleus (see online
Supplementary Materials for details).  The $P_{cm}$ distributions,
that describe the influence of the $A-2$ nuclear system on the SRC
pairs, can also be obtained from mean-field
calculations~\cite{CiofidegliAtti:1995qe,Colle:2013nna}.  These
calculated $P_{cm}$ distributions are 
consistent with the experimentally extracted
ones~\cite{Cohen:2018gzh}.  We therefore do not expect them to have
significant scale- and scheme-dependence.

We used Glauber-based calculations to estimate the model- and
kinematics-dependent SCX corrections~\cite{Colle:2015lyl}. 
We applied these corrections in two ways, 
to correct the GCF cross-section calculations and compare them to the uncorrected data,
and also to correct our data in order to directly extract the relative
abundance of $pp$- and $np$-SRC pairs.
As the Glauber calculations describe the influence of the $A-2$ system on the
measured reactions, and are based on measured NN scattering cross-sections,
we also do not expect them to have significant scale- and scheme-dependence.

\begin{figure} [t]
\includegraphics[width=\columnwidth, height=5.7cm]{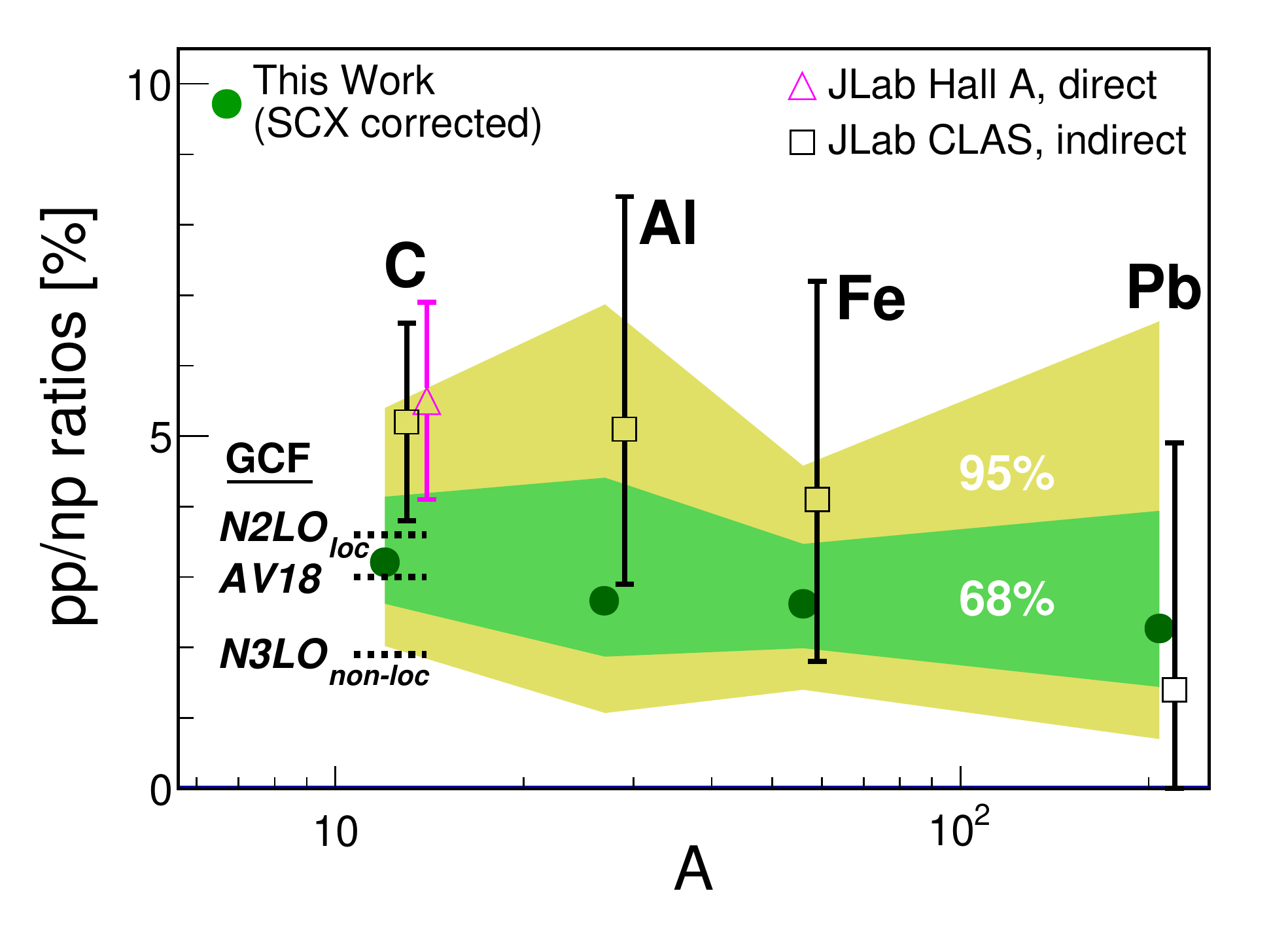}
\caption{\label{fig:1} (color online) Extracted ratios of $pp$- to
$np$-SRC pairs plotted versus atomic weight $A$. The filled green circles show the ratios of $pp$- to
$np$-SRC pairs extracted from \eepp / \eepn cross-section ratios corrected for 
SCX using Eq.~\ref{eq:1}. The
shaded regions mark the 68\% and 95\% confidence limits on the
extraction due to uncertainties in the measured cross-section ratios
and SCX correction factors (see online supplementary materials for details).
The magenta triangle shows the carbon data of~\cite{subedi08}, 
which were also corrected for SCX. 
The open black squares show the indirect extraction of Ref. \cite{hen14}.
The uncertainties on both previous extractions mark the 68\% (i.e. 1$\sigma$) confidence limits.
The horizontal dashed lines show the $^{12}$C GCF-calculated contact ratios for different $NN$ potentials using contact values fitted directly to the measured cross-section ratios.
See text for details.}
\end{figure}

% discussion of Precoil dependence comparisons with GCF
Figure~\ref{fig:0} shows the measured reduced cross section ratios
(without SCX corrections) compared with SCX-corrected GCF cross-section
ratio calculations. The GCF calculations are done for $^{12}$C,
following Ref.~\cite{Weiss:2018tbu} using three NN potentials: the phenomenological
AV18~\cite{wiringa95}, a chiral EFT local
N2LO(1.0)~\cite{Gezerlis:2013ipa,Gezerlis:2014zia}, and a chiral EFT non-local
N3LO(600)~\cite{Epelbaum:2008ga}.
The uncertainties in the calculation include contributions from the
measured width of the SRC pair cm motion ($\sigma_{cm} = 150 \pm 20$
MeV/c)~\cite{Cohen:2018gzh}, the residual $A-2$ excitation energy ($E^* = 0$
to $30$ MeV), SCX probabilities (see table I in the online supplementary materials),
values of the contact terms, and off-shell
electron-nucleon cross-section model ($\sigma_{CC1}$ and
$\sigma_{CC2}$ from Ref.~\cite{deforest83}, using the form factor parameterization of Ref.~\cite{Kelly:2004hm}). 

We calculated the cross section ratios for two different sets of
$pp$ (spin-0) to $np$ (spin-1 only) contact ratios.
The `Old' ones used those previously determined in
Ref.~\cite{Weiss:2018tbu} while the `New' ones used contacts directly fitted to the
new data presented here.
See online supplementary materials for details.

The previously determined~\cite{Weiss:2018tbu} $pp$ to $np$ contact
ratios for the AV18, local N2LO, and non-local N3LO interactions are:
$7.1\% \pm 1.5\%$; $5.2\% \pm 1.1\%$; and $4.0\% \pm 0.8\%$
respectively. The contact values fitted to this data are significantly
lower: $3.0\% \pm 0.8\%$; $3.6\% \pm 1.0\%$; and $1.9\% \pm 0.5\%$ for
the three different potentials.  A large part of this reduction
(factor of about 1.7) is due to the more complete SCX corrections
applied here, as compared to that available ten years ago
\cite{frieds65} for the data used in Ref.~\cite{Weiss:2018tbu}.

The fact that the same cross-section ratios are obtained from GCF calculations using combinations of different NN interactions and  contact ratios and shows the importance of preforming data-theory conparisons at the cross-section level, accounting for the conplete integral over the SRC pairs relative and c.m. momentum distributions in the extraction of the nuclear contacts~\cite{Weiss:2018tbu}.

Figure~\ref{fig:1} shows the alternative approach where we directly
correct the data for SCX effects. This allows determining
the $pp$ to $np$ fraction with different and somewhat simplified
assumptions than those used by the GCF calculation. For this we
consider all recoil proton momenta and express the relative abundance
of $pp$- to $np$-SRC pairs as (see derivation in the online supplementary materials):
\begin{equation}
\frac{\#pp\hbox{-}SRC}{\#np\hbox{-}SRC} = \frac{1}{2} \frac{2 R P^{np}_{A} - P^{[n]p}_{A} - P^{p[n]}_{A}/\sigma_{p/n}}{P^{pp}_{A}-2\sigma_{p/n} R P^{[p]p}_{A} - 2 R \eta_A P^{n[n]}_{A}},
\label{eq:1}
\end{equation}
where $\eta_A = \frac{\#nn\hbox{-}SRC}{\#pp\hbox{-}SRC}$ and $R$ is
reduced cross section ratio of Eq.~\ref{eq:R}.
$P^{NN}_{A}$ is the probability for scattering off an $NN$ pair without subsequent SCX,
and $P^{[N]N}_{A}$ and $P^{N[N]}_{A}$ are the probabilities for
scattering off an $NN$ pair and having either the leading or recoil
nucleon undergo SCX, respectively. The values and uncertainties of the
parameters used in Eq.~\ref{eq:1} are listed in table I of the online supplementary materials. 

While the current analysis uses the SCX calculations of
Ref.~\cite{Colle:2015lyl} and the formalism detailed in the online supplementary materials,
other calculations for these corrections can be applied in the future.
See online supplementary materials for detailes on the numerials evaluation of Eq.~\ref{eq:1} and its uncertainty.

These SCX-corrected $pp/pn$ ratios agree within uncertainty with the
ratios previously extracted from \Aeepp and \Aeep events \cite{hen14},
which assumed that all
high-missing momentum nucleons belong to SRC pairs.
In addition, the SCX-corrected $pp / np$ ratio is in better agreement
with the GCF contacts fitted here but is not inconsistent with those
determined in Ref.~\cite{Weiss:2018tbu}. This is a significant
achievement of the GCF calculations that opens the way for detailed
data-theory comparisons. This will be possible using future higher
statistics data that will allow finer binning in both recoil and
missing momenta.

The $pp/np$ ratios measured directly in this work are somewhat lower than
both previous indirect measurements on nuclei from C to Pb\cite{hen14}, 
and previous direct measurements on C \cite{subedi08}.
This is due to the more sophisticated SCX calculations used
in this work \cite{Colle:2015lyl} compared to the previous ones
\cite{frieds65}. This is consistent with the lower values of the $pp$
to $np$ contact extracted from GCF calculations fit to this data
mentioned above.

To conclude, we report the first measurements of high
momentum-transfer hard exclusive $np$ and $pp$ SRC pair knockout
reactions off symmetric ($^{12}$C) and medium and heavy neutron-rich
nuclei ($^{27}$Al, $^{56}$Fe, and $^{208}$Pb).  We find that the
reduced cross-section ratio for proton-proton to proton-neutron
knockout equals $\sim 6\%$, consistent with previous measurements off
symmetric nuclei. Using model-dependent SCX corrections, we also
extracted the relative abundance of $pp$- to $pn$-SRC pairs in the
measured nuclei. As expected, these corrections reduce the
$pp$-to-$np$ ratios to about $3\%$, so that the measured reduced
cross-section ratios are an
upper limit on the relative SRC pairs abundance ratios.

The data also shows good agreement with GCF calculations using
phenomenological as well as local and non-local chiral $NN$
interactions, allowing for a higher precision determination of nuclear
contact ratios and a study of their scale- and scheme-dependence.
While the contact-term ratios extracted for phenomenological and
local-Chiral interactions are consistent with each other, they are
larger than those obtained for the non-local Chiral interaction
examined here. Forthcoming data with improved statistics will allow
mapping the missing and recoil momentum dependence of the measured
ratios. This will facilitate detailed studies of the origin,
implications, and significance of such differences.

Previous work~\cite{hen14} measured \Aeep and \Aeepp events and
derived the relative probabilities of $np$ and $pp$ pairs assuming
that all high-missing momentum \Aeep events were due to scattering
from SRC pairs. The agreement between the $pp$/$np$ ratios directly
measured here and those of the previous indirect measurement, as well
as with the factorized GCF calculations, strengthens the $np$-pair
dominance theory and also lends credence to the previous assumption
that almost all high-initial-momentum protons belong to SRC pairs in
nuclei from C to Pb.

\begin{acknowledgments}
We acknowledge the efforts of the staff of the Accelerator and Physics Divisions at Jefferson Lab that made this experiment possible. We are also grateful for many fruitful discussions with L.L. Frankfurt, M. Strikman, J. Ryckebusch, W. Cosyn, M. Sargsyan, and C. Ciofi degli Atti. The analysis presented here was carried out as part of the Jefferson Lab Hall B Data-Mining project supported by the U.S. Department of Energy (DOE). The research was supported also by the National Science Foundation, the Pazy Foundation, the Israel Science Foundation, the Chilean Comisión Nacional de Investigación Científica y Tecnológica, the French Centre National de la Recherche Scientifique and Commissariat a l'Energie Atomique the French-American Cultural Exchange, the Italian Istituto Nazionale di Fisica Nucleare, the National Research Foundation of Korea, and the UK's Science and Technology Facilities Council. Jefferson Science Associates operates the Thomas Jefferson National Accelerator Facility for the DOE, Office of Science, Office of Nuclear Physics under contract DE-AC05-06OR23177. The raw data from this experiment are archived in Jefferson Lab's mass storage silo.
\end{acknowledgments}

\bibliography{npTriple_bib}

\clearpage

\appendix

%================================================================
%App. for SCX
%================================================================
\section{Supplementary Materials}

\subsection{Formalism}
In the absence of FSI, assuming scattering from an SRC pair, the $A(e,e'np)$ and $A(e,e'pp)$ measured cross-sections can be written as:

\begin{equation}
\begin{split}
{A(e,e'pp)} &\propto \#pp_{A}\cdot 2\cdot\sigma_{ep}, \\
{A(e,e'np)} &\propto \#np_{A}\cdot\sigma_{en},
\end{split}
\label{eq:2}
\end{equation}
where $\#pp_{A}$ ($\#np_{A}$) is the number of proton-proton (neutron-proton) pairs in nucleus $A$ and $\sigma_{ep}$ ($\sigma_{en}$) is the electron-proton (electron-neutron) cross-section. 

With FSI, one should take into account contributions from all $NN$-SRC pairs that can lead to the same measured final state, the effects of nuclear transparency and SCX. 

Using the notation defined in the main text for the SCX probabilities, Eq.~\ref{eq:2} can be extended as:
\begin{equation}
\begin{split}
{A(e,e'pp)}\propto & \#pp_{A}\cdot 2\sigma_{ep} \cdot P_{A}^{pp} \cdot T_{A,pp}+ \\
& \#np_{A}\cdot\sigma_{en} \cdot p_{A}^{[n]p} \cdot T_{A}^{*}+ \\ 
& \#pn_{A}\cdot\sigma_{ep} \cdot P_{A}^{p[n]} \cdot T_{A}^{*}, \\ \\
{A(e,e'np)}\propto & \#np_{A}\cdot\sigma_{en}\cdot P_{A}^{np}\cdot T_{A,np}+ \\
&\#pp_{A}\cdot2\sigma_{ep}\cdot P_{A}^{[p]p}\cdot T_{A}^{*}+ \\ 
& \#nn_{A}\cdot 2\sigma_{en}\cdot P_{A}^{n[n]}\cdot T_{A}^{*},
\end{split}
\label{eq:3}
\end{equation}
where $T_{pp}$ ($T_{np}$) is the nuclear transparency for two protons (neutron-proton) and $T^{*}$ is the transparency associated with a SCX process.

Eq. 2 in the main text can be obtained from Eq.~\ref{eq:3} above by forming the $A\eepp$ / $A\eepn$ ratio and assuming that $T^{*}=\frac{1}{2}(T_{pp}+T_{np})=T_{pp}=T_{np}$. The latter approximation is valid when considering high-$Q^2$ reactions with a high energy leading proton/neutron that has the same nuclear transparency for $pp$ and $np$ pairs~\cite{Ryckebusch:2003fc,Colle:2015lyl}.

Following Ref.~\cite{Weiss:2018tbu}, and adopting the kinematical notations of Fig. 1 of the main text, the GCF calculated cross-section for the $A(e,e'NN)$ process without FSI can be expressed as:
\begin{equation}
\begin{split}
& \frac{d^8\sigma}{dQ^2 dx_B d\phi_e d^3\vec{p}_{CM} d\Omega_{rec}} = \\
& \frac{\sigma_{eN}}{32\pi^4}  \times n(\vec{p}_{CM}) \times \sum\limits_\alpha C_\alpha|\tilde{\varphi}^\alpha(|\vec{p}_{CM} - 2\vec{p}_{rec}|)|^2  \times \\
& \frac{E_N E_{rec} p_{rec}^2}{|E_{rec}(p_{rec} -Z\cos\theta_{Z,rec}) + E_N p_{rec}|}   \times
\frac{\omega}{2E_{beam} E_e x_B} ,
\end{split}
\label{eq:xsection_gcf}
\end{equation}
where the subscript `N' (`rec')  stands for the leading (recoil) nucleon, `e' stands for the scattered electron, $\vec{Z} = \vec{q} + \vec{p}_{CM}$ and $\theta_{Z,rec}$ is the angle between $\vec{Z}$ and $\vec{p}_{rec}$.

The second line shows the GCF factorization
to an off-shell electron-nucleon cross section $\sigma_{eN}$ (taken from Ref.~\cite{deforest83}, 
using the form factor parameterization of Ref.~\cite{Kelly:2004hm}),
times an SRC pair center-of-mass momentum distribution 
$n(\vec{p}_{CM})$ (given by a three-dimensional Gaussian with width of $150 \pm 20$ MeV/c~\cite{Cohen:2018gzh,CiofidegliAtti:1995qe,Colle:2013nna}), 
times a summation over all SRC pairs that can contribute to a given NN final state (without SCX).
The index $\alpha$ represents the quantum numbers of an SRC pair such that for $(e,e'pp)$ $\alpha$ stands for spin-0 pp pairs 
while for $(e,e'pn)$ it is summed over both spin-0 and spin-1 pn pairs. 
The functions $\tilde{\varphi}^\alpha$ are the universal SRC pairs relative momentum distributions obtained by solving the zero-energy two-body Schrodinger equation of a NN pair in quantum state $\alpha$ using a given NN potential model. 
$C_\alpha$ are nuclear contact terms that determine the relative abundance of SRC pairs with a given quantum state.
For pp (nn) pairs $C_\alpha$ is equal to twice the nuclear contact for pp-SRC (nn-SRC) pairs.
The third line is simply a Jacobian term.

The total measured two-nucleon knockout cross-section, $\sigma^{GCF}_{A(e,e'NN)}$, is obtained by integrating Eq.~\ref{eq:xsection_gcf} over the experimental acceptance as detailed below.

FSI can be introduced to the GCF calculation in a similar manner as in Eq.~\ref{eq:3} using:
\begin{equation} 
\begin{split}
\sigma^{Exp}_{A(e,e'pp)} = &\sigma^{GCF}_{A(e,e'pp)} \cdot P_{A}^{pp}\cdot T_{A,pp} + \\
& \sigma^{GCF}_{A(e,e'np)}\cdot p_{A}^{[n]p}\cdot T_{A}^{*} + \\
& \sigma^{GCF}_{A(e,e'pn)}\cdot P_{A}^{p[n]}\cdot T_{A}^{*}, \\ \\
\sigma^{Exp}_{A(e,e'np)} = & \sigma^{GCF}_{A(e,e'np)} \cdot P_{A}^{np}\cdot T_{A,np} + \\
& \sigma^{GCF}_{A(e,e'pp)} \cdot P_{A}^{[p]p}\cdot T_{A}^{*} + \\
&\sigma^{GCF}_{A(e,e'nn)}\cdot P_{A}^{n[n]}\cdot T_{A}^{*}.
\end{split}
\label{eq:scx_gcf}
\end{equation}

\subsection{Calculation details}
The evaluation of Eq. 2 in the main text, and the estimation of its uncertainties, was done
following~\cite{hen14}, using a Monte-Carlo technique where its PDF was extracted
from repeated calculations using different input values. In each calculation the
values of the different parameters (experimental cross-section, SCX probabilities
etc.) were randomly chosen from a Gaussian distribution centered at the measured 
or calculated value with width ($1\sigma$) that equaled their associated uncertainties.
The cross-section ratios, $R$, are listed in Table II of the main text, the SCX probabilities
are listed in Table~\ref{tab:3} below (based on the calculations of Ref.~\cite{Colle:2015lyl}).
Notice that all probabilities (for protons, neutrons, leading, and recoil) increase with
$A$. The probabilities for the lower-momentum recoil nucleons are larger than those for
the leading nucleons. The SCX probabilities for protons are larger than for neutrons.
For the kinematics of the current measurement $\sigma_{p/n}=\frac{\sigma_{ep}}{\sigma_{en}}
=2.30\pm0.15$. For asymmetric nuclei $\eta_{A}=\frac{\#nn_{A}}{\#pp_{A}}$ was drawn
from a uniform distribution between unity and the combinatorial ratio of possible 
$nn$ and $pp$ pairs in a given asymmetric nucleus. The resulting FSI-corrected 
$pp/np$ SRC pairs ratio, presented in Fig. 3 of  the main text, show the most-probable
value, with confidence bands enclosing 68\% of the PDF.

The GCF results shown in Fig. 2 of the main text were obtained by using a Monte-Carlo event 
generator to integrate Eq.~\ref{eq:xsection_gcf} over the experimental acceptance by
considering only kinematics where the scattered electron, knockout nucleon and recoil
proton would all have been detected in CLAS (i.e., the momentum vectors all point to active regions in CLAS), and furthermore pass the
event selections cuts listed in Table I of the main text. 
Transparency and SCX effects were 
introduced to the integrated cross-section using Eq.~\ref{eq:scx_gcf}. The uncertainty
on the calculation was estimated by repeating the calculation many times while varying
the input parameters to Eq.~\ref{eq:scx_gcf} and~\ref{eq:xsection_gcf} by their 
uncertainties detailed in the text, and extracting the 68\% and 95\% confidence intervals
of the calculation due to these variations, similar to the procedure described above for Eq. 2 of the main text.

\begin{table}[H]
\caption{\label{tab:3} The SCX probabilities for different $NN$ pairs
and nuclei. The nucleons in brackets underwent a SCX interaction,
the ones not in brackets did not. Thus, $P^{[p]n}$ indicates the
probability that the proton in a knocked-out $pn$ pair undergoes a
SCX reaction and the neutron does not.}
\begin{tabular}{|ccccc|}
\hline 
& C & Al & Fe & Pb \\
\hline 
\hline 
$P^{pp}$ & \footnotesize{$0.908\pm0.006$} & \footnotesize{$0.897\pm0.009$} & \footnotesize{$0.891\pm0.010$} & \footnotesize{$0.860\pm0.013$} \\
$P^{[p]p}$ & \footnotesize{$0.041\pm0.003$} & \footnotesize{$0.046\pm0.004$} & \footnotesize{$0.048\pm0.005$} & \footnotesize{$0.059\pm0.006$} \\ 
$P^{p[p]}$ & \footnotesize{$0.048\pm0.003$} & \footnotesize{$0.054\pm0.005$} & \footnotesize{$0.057\pm0.006$} & \footnotesize{$0.074\pm0.007$} \\ 
$P^{[pp]}$ & \footnotesize{$0.003\pm0.0002$} & \footnotesize{$0.004\pm0.0003$} & \footnotesize{$0.004\pm0.0003$} & \footnotesize{$0.007\pm0.0006$} \\
$P^{p[n]}$ & \footnotesize{$0.041\pm0.003$} & \footnotesize{$0.047\pm0.005$} & \footnotesize{$0.047\pm0.005$} & \footnotesize{$0.047\pm0.005$} \\ 
$P^{[p]n}$ & \footnotesize{$0.035\pm0.002$} & \footnotesize{$0.043\pm0.004$} & \footnotesize{$0.046\pm0.005$} & \footnotesize{$0.061\pm0.006$} \\ 
$P^{np}$ & \footnotesize{$0.922\pm0.005$} & \footnotesize{$0.907\pm0.008$} & \footnotesize{$0.903\pm0.009$} & \footnotesize{$0.887\pm0.010$} \\
$P^{[n]p}$ & \footnotesize{$0.035\pm0.002$} & \footnotesize{$0.040\pm0.004$} & \footnotesize{$0.040\pm0.004$} & \footnotesize{$0.040\pm0.004$} \\ 
$P^{n[p]}$ & \footnotesize{$0.041\pm0.003$} & \footnotesize{$0.051\pm0.005$} & \footnotesize{$0.054\pm0.006$} & \footnotesize{$0.072\pm0.008$} \\ 
$P^{[np]}$ & \footnotesize{$0.002\pm0.0001$} & \footnotesize{$0.003\pm0.0002$} & \footnotesize{$0.004\pm0.0003$} & \footnotesize{$0.005\pm0.0004$} \\
$P^{n[n]}$ & \footnotesize{$0.048\pm0.003$} & \footnotesize{$0.050\pm0.005$} & \footnotesize{$0.049\pm0.005$} & \footnotesize{$0.048\pm0.005$} \\ 
\hline 
\end{tabular}
\end{table}

\end{document}